\documentclass[aps,prl,notitlepage,twocolumn]{revtex4-1}
\usepackage{amsmath,amssymb}
\usepackage{dsfont}
\usepackage{graphicx}

\begin{document}

\newcommand{\bra}[1]    {\left\langle #1\right|}
\newcommand{\ket}[1]    {\left| #1 \right\rangle}
\newcommand{\tr}[1]    {{\rm Tr}\left[ #1 \right]}
\newcommand{\av}[1]    {\left\langle #1 \right\rangle}
\newcommand{\PRL}[1]{Phys. Rev. Lett.~\textbf{#1}}
\newcommand{\PRA}[1]{Phys. Rev.~A~\textbf{#1}}
\newcommand{\proj}[1]{\ket{#1}\bra{#1}}

\author{T. Wasak$^1$, P. Sza\'nkowski$^1$, P. Zi\'n$^2$, M. Trippenbach$^1$ and J. Chwede\'nczuk$^1$}
\affiliation{
  $^1$Faculty of Physics, University of Warsaw, ul.\ Ho\.{z}a 69, PL--00--681 Warszawa, Poland\\
  $^2$National Centre for Nuclear Research, ul. Ho\.za 69, PL-00-681 Warsaw, Poland}

\title{Cauchy-Schwarz inequality and particle entanglement}

\begin{abstract}
  The Glauber-Sudarshan P-representation is used in quantum optics to distinguish between semi-classical and genuinely quantum electromagnetic fields.
  We employ the analog of the P-representation to systems of identical bosons and show that the violation of the Cauchy-Schwarz inequality for the second-order correlation function
  is a proof of particle entanglement.
  The present derivation applies to any quantum system of identical bosons, with either fixed or fluctuating number of particles, provided that there is no coherence between different number states.
  In the light of recent experimental advances in single-particle detection,
  the violation of the Cauchy-Schwarz inequality may become an easily accessible entanglement probe in correlated many-body systems.
\end{abstract}
\maketitle

Although the foundations of quantum and classical physics are much different, it is often difficult to construct a simple criterion of ``quantumness'' of a particular system.
A good example of a non-classical behavior is, according to the Schr\"odinger equation, the ability of particles to exist in superpositions of quantum states.
The most prominent manifestation of such superposition
is the Young double-slit experiment for massive particles, which confirms their wave character.
For optical waves it is the opposite -- the non-classical electromagnetic field is that consisting of individual photons. The challenging question whether, and in what sense,
the pulse of light is quantum, was among the key issues triggering the development of quantum optics. 

The problem was formalized by Glauber and Sudarshan in their studies on coherence in the context of correlation functions \cite{sud,glaub}. 
They employed the coherent states $\ket{\Phi}$ defined by the relation $\hat{\mathcal E}^{(+)}(x)\ket\Phi=\Phi(x)\ket\Phi$, where $\hat{\mathcal E}^{(+)}(x)$ is
the positive-frequency part of the electromagnetic field $\hat{\mathcal E}(x)$, and expressed the density matrix using the so-called P-representation
\begin{equation}\label{em}
  \hat\rho=\int\!\!\mathcal{D}\Phi\ket{\Phi}\!\!\bra{\Phi}\mathcal{P}(\Phi).
\end{equation}
The symbol $\mathcal{D}\Phi$ denotes the integration measure over the set of complex fields $\Phi$.
The state of light is classical if the outcome of the measurement can be explained in terms of classical electromagnetic fields, which happens when the P-representation can be interpreted as a probability 
distribution, which means that it is normalized and
\begin{equation}\label{cond_em}
  \int\limits_{\mathcal V}\!\mathcal{D}\Phi\,\mathcal{P}(\Phi)\geqslant0
\end{equation}
for {\it any} volume $\mathcal V$. When the P-representation does not satisfy condition (\ref{cond_em}), the field is said to be quantum.

Once the electromagnetic field is quantized, photons can be treated on a more equal foot with other particles.
It is then reasonable to ask the question about correlations between individual particles and
in this context the concept of particle entanglement emerges \cite{ent1,ent2}. The possibility for particles to be entangled, which is a purely quantum phenomenon, has rather dramatic consequences.
The quantumness of entanglement is underlined by the word  ``paradox'' often
used to describe some highly counter-intuitive phenomena such as the Einstein-Podolsky-Rosen (EPR) paradox \cite{epr} and the related Schr\"odinger's cat problem.
Apart from fundamental aspects, systems of entangled particles have applications in quantum information \cite{quantinf}, teleportation \cite{wootters1,wootters2}
or ultra-precise metrology \cite{giovanetti,pezze}.
Entanglement is also believed to contribute to the extreme efficiency of the energy transfer in the process of photosynthesis \cite{synt}.

Much as a fascinating consequence of quantum mechanics, entanglement is also elusive. It is not simple to entangle particles on demand,
because this requires complicated experimental strategies and it is difficult
to protect them from the destructive influence of the environment, which inevitably leads to decoherence \cite{decoh1,decoh2,decoh3,decoh4,decoh5}.
Finally, even if a non-classical state reaches detectors fairly intact, it is often not clear
which quantity should be measured to witness entanglement.

This last difficulty is related to the very definition, which states that a particle-entangled state is such that {\it is not} separable, meaning
that it cannot be written as a mixture of product states of $N$ particles \cite{sor, peres, wer}
\begin{equation}\label{sep}
    \hat\rho_{\rm sep}=\sum_i\,p_i\,\hat\rho^{(1)}_i\otimes\ldots\otimes\hat\rho^{(N)}_i.
\end{equation}
Here $p_i$'s are non-negative weights that add up to unity. The consequence of this indirect definition is that to characterize entanglement we usually first  refer to some
bounds achievable by separable states.
A good example is the two-mode quantum interferometer, which utilizes a collection of $N$ qubits in state $\hat\rho$ to determine an unknown phase $\theta$.
If the precision of the parameter estimation $\Delta\theta$ is better than shot-noise $\Delta\theta=N^{-\frac12}$ (the smallest error attainable with separable states), then $\hat\rho$ is entangled \cite{pezze}.
However, in most cases the argument cannot be reversed, just because we do not know what {\it is} the entangled state.

We underline that  entanglement of identical bosons might simply result from the symmetrization. In context of quantum information, such type of entanglement has been disregarded
because in most protocols the particles are well-separated and can be addressed individually by local measurements \cite{sym1,sym2,sym3,sym4}. Nevertheless, it has been recently demonstrated that
entanglement of identical particles can be mapped onto the mode entanglement useful for quantum information by means of simple operations like the mode splitting \cite{plenio}. 
As an illustration, consider a pure state $\ket{1,1}$ of a pair of identical bosons occupying far separated modes. This state is 
particle-entangled due to symmetrization, but naturally it is not mode entangled. However, if the particles are brought together and simultaneously pass a beam-splitter, the state is transformed 
into the NOON state $\frac1{\sqrt2}(\ket{2,0}+\ket{0,2})$, which is both mode- and particle-entangled. This demonstrated that using a beam-splitter it
is possible to extract useful correlations between the modes starting from a state which is only particle-entangled.

The results of \cite{plenio} shed new light onto the entanglement of identical bosons, which so far has been regarded mainly as a resource for sub shot-noise quantum metrology. As mentioned above, 
the precision of parameter estimation can be improved if the input port of an interferometer is fed with a spin-squeezed state \cite{wine,kita,ita1}, which is entangled due to the indistinguishability 
of the constituent bosons \cite{sor}.
The spin-squeezing was recently generated with cold bosonic atoms \cite{esteve,mzi_vienna,app,gross,riedel,vule,chen} and its usefulness for ultra-precise interferometry was demonstrated.

In this work we employ the analog of the P-representation (\ref{em}) to formulate a new yet simple criterion for particle entanglement.
It is based on the measurements of the second order correlation function and is valid for any system of indistinguishable bosons with either fixed or fluctuating number of particles, 
for as long as coherences between different number states are absent. 
%% In this work we show that the analog of the P-representation allows to formulate a simple particle entanglement criterion based on the
%% second order correlation function of identical  bosons.
The discussion begins by stating that a separable pure state of $N$ identical bosons $\ket{\phi;N}$ must be a product of $N$ identical single-particle orbitals $\ket\phi$, i.e.
\begin{equation}\label{coh_def}
  \ket{\phi;N}=\ket{\phi}^{\otimes N}.
\end{equation}
%If orbitals were different, the state would have to be symmetrized to preserve indistinguishability of bosons. However, symmetrized states other then (\ref{coh_def}) are no longer separable.
%If orbitals were different, then symmetrization would introduce entanglement between the particles so the state would not be separable anymore.
When the bosonic field operator $\hat\Psi(x)$ acts on the state (\ref{coh_def}), the result is
\begin{equation}\label{coh}
  \hat\Psi(x)\ket{\phi;N}=\sqrt N\phi(x)\ket{\phi;N-1},
\end{equation}
which is a fixed-$N$ counterpart of the property of a coherent state of light $\ket\Phi$.
In Eq.~(\ref{coh}), $\phi(x)$ is a single-particle function, which determines the spatial properties of the system.

As we prove in the Supplementary Materials, rephrasing the arguments of \cite{proof1,proof2}, the general separable state of $N$ identical bosons is a mixture of different  states (\ref{coh_def}),
\begin{equation}\label{den}
  \hat\rho=\int\!\!\mathcal{D}\phi\,\ket{\phi;N}\!\!\bra{\phi;N}\mathcal{P}(\phi),
\end{equation}
where $\mathcal{D}\phi$ denotes the integration over the complex field $\phi$ and $\mathcal{P}(\phi)$ is the probability distribution, which means that it is normalized and its integral
over any volume $\mathcal V$ is non-negative, i.e.
\begin{equation}\label{cond_N}
  \int\limits_{\mathcal V}\!\mathcal{D}\phi\,\mathcal{P}(\phi)\geqslant0.
\end{equation}
There is a direct analogy between the P-representation from Eq.~(\ref{em}) and $\mathcal{P}(\phi)$ from Eq.~(\ref{den}).
Recall that if the former does not satisfy condition (\ref{cond_em}), then the electromagnetic field is genuinely quantum.
Analogically for $N$ indistinguishable bosons, if condition (\ref{cond_N}) is not fulfilled, then the density matrix cannot be written as a statistical mixture (\ref{den}) meaning that particles
are entangled.

\begin{figure}[hbt]
  \centering
  \includegraphics[clip,scale=.16]{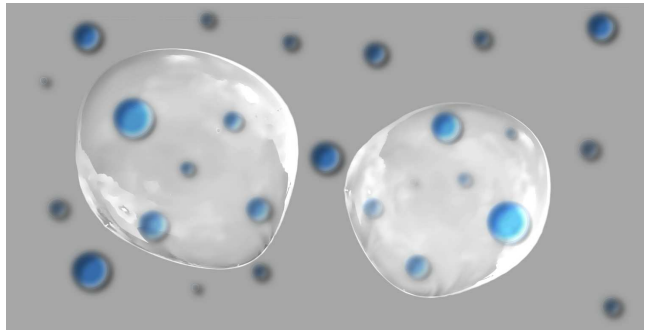}
  \caption{Schematic illustration of the system of $N$ identical bosons with two designated regions. By means of the coefficient $\mathcal C$ defined in Eq.~(\ref{csi}), 
    the Cauchy-Schwarz inequality quantifies the strength of the second order correlations between the particles occupying these two regions. 
    All values of $\mathcal C$, which exceed unity, are not allowed by classical mechanics and signfy particle entanglement.}
  \label{scales}
\end{figure}

In the domain of quantum optics, various criteria were introduced to verify if the state's P-representation satisfies condition (\ref{cond_em}). 
Among them we find the Cauchy-Schwarz inequality (CSI) for the second order correlation function and the related number squeezing between two regions \cite{shch,hillery,tests}.
For instance, the CSI can indicate if the P-representation is partially negative, which signifies entanglement in a two-party system \cite{wodk}. 
Here we show to which extent these criteria apply to systems of $N$ identical bosons as probes of particle entanglement.
%Below we show that analogically, the CSI is a criterion for particle entanglement in a system of $N$ identical bosons.

For separable states (\ref{den}), the second-order correlation function, calculated using Eq.~(\ref{coh}) is
\begin{eqnarray}
  &&\mathcal{G}^{(2)}(x,x')=\left\langle\hat\Psi^\dagger(x)\hat\Psi^\dagger(x')\hat\Psi(x')\hat\Psi(x)\right\rangle\nonumber\\
  &&=N(N-1)\int\!\!\mathcal{D}\phi\,\mathcal{P}(\phi)|\phi(x)|^2|\phi(x')|^2\label{g2}.
\end{eqnarray}
For the considerations that follow, we introduce two regions $a$ and $b$ having volumes $V_a$ and $V_b$ and an integrated second-order correlation function (\ref{g2})
%% \begin{eqnarray}\label{integrated}
%%   &&\mathcal{G}^{(2)}_{ij}=\int\limits_{V_i}\!\! dx\!\!\int\limits_{V_j}\!\! dx'\mathcal{G}^{(2)}(x,x')\nonumber\\
%%   &&=N(N-1)\int\!\!\mathcal{D}\phi\,\mathcal{P}(\phi)I_i(\phi)I_j(\phi)
%% \end{eqnarray}
\begin{equation}\label{integrated}
  \mathcal{G}^{(2)}_{ij}\!\!=\!\!\int\limits_{V_i}\!\! dx\!\!\int\limits_{V_j}\!\! dx'\mathcal{G}^{(2)}(x,x')\!=\!N(N-1)\!\!\int\!\!\mathcal{D}\phi\,\mathcal{P}(\phi)I_i(\phi)I_j(\phi)
\end{equation}
where $I_{i/j}(\phi)=\int_{V_{i/j}}\!\! dx|\phi(x)|^2$ with $i$ and $j$ being either $a$ or $b$. We now apply the Cauchy-Schwarz inequality (CSI)
for non-negative functions $f_{a/b}(\phi)=\sqrt{\mathcal{P}(\phi)}\,I_{a/b}$,
\begin{equation}
 \int\!\!\mathcal{D}\phi\,f_a(\phi)\,f_b(\phi)\leqslant\sqrt{\int\!\!\mathcal{D}\phi f^2_a(\phi)}\sqrt{\int\!\!\mathcal{D}\phi f^2_b(\phi)},
\end{equation}
which gives
\begin{equation}\label{csi}
  \mathcal{C}\equiv\frac{\mathcal{G}^{(2)}_{ab}}{\sqrt{\mathcal{G}^{(2)}_{aa}\mathcal{G}^{(2)}_{bb}}}\leqslant1.
\end{equation}
On the other hand, a particle-entangled twin-Fock state of
$N$ identical bosons equally occupying two modes, i.e. $\ket{\psi_{\rm tf}}=\ket{\frac N2,\frac N2}$, gives $\mathcal{C}=1+\frac2{N-2}$, which violates the CSI.
To summarize, we have shown that (a) the CSI is satisfied by all separable states of identical bosons and (b) there exists an entangled state, by which the CSI is violated.
Therefore the CSI can be treated as a criterion for particle entanglement.

A broad family of entangled states, which violate the CSI can be identified using the number-squeezing parameter defined as a variance of the population imbalance operator between the two regions,
$\eta^2=\frac{\av{\hat n^2}-\av{\hat n}^2}{n_{\rm tot}}$. Here $\hat n=\hat n_a-\hat n_b$ and $n_{\rm tot}=\av{\hat n_a}+\av{\hat n_b}$ with
$\hat n_{a/b}=\int_{V_{a/b}}\!\! dx\,\hat\Psi^\dagger(x)\hat\Psi(x)$. This parameter can be expressed in terms of the local- and
cross-correlations as follows
%% \begin{equation}
%%   \eta^2=N_a+N_b-(N_a-N_b)^2+\mathcal{G}^{(2)}_{aa}+\mathcal{G}^{(2)}_{bb}-2\mathcal{C}\sqrt{\mathcal{G}^{(2)}_{aa}\mathcal{G}^{(2)}_{bb}},
%% \end{equation}
\begin{equation}\label{ns}
  \eta^2=1+\frac{\mathcal{G}^{(2)}_{aa}+\mathcal{G}^{(2)}_{bb}-2\mathcal{G}^{(2)}_{ab}-\av{\hat n}^2}{n_{\rm tot}}
\end{equation}
First, consider a balanced state with $\av{\hat n}=0$ and $\mathcal{G}^{(2)}_{aa}=\mathcal{G}^{(2)}_{bb}$.
In this case $\eta^2=1-2\frac{(1-\mathcal{C})\,\mathcal{G}^{(2)}_{aa}}{n_{\rm tot}}$. The system is number squeezed, $\eta^2<1$ when  $\mathcal{C}>1$,
so the number-squeezing is equivalent to the violation of the CSI and signifies particle entanglement. If the state
is not balanced, one cannot link the number squeezing with the CSI, due to presence of the non-vanishing term $\av{\hat n}^2$.
As an example, take a separable state of $N$ particles in a pure state $\ket{\phi_0}^{\otimes N}$, divided into two unequal parts. We obtain that $\eta^2=1-\frac1N\av{\hat n}^2<1$,
so the number squeezing is present without entanglement.
%% This is in sharp contrast to quantum optics, where the squeezing of light is related to the violation of conditions $P$-representation of squeezed states of light violates conditions (\ref{cond_em}). The discrepancy between these two cases arises from
%% the difference between the state (\ref{coh_def}), which has a fixed number of particles and the coherent state of light. While in the former case, the
Clearly, the CSI is more
universal then the number squeezing parameter, because its violation always implies entanglement of identical bosons and its construction does not require any assumptions about the two regions.
Nevertheless, the relation between the integrated correlation functions
(\ref{integrated}) and $\eta^2$ from Eq.~(\ref{ns}) is a strong suggestion that the violation of the CSI is more likely to manifest in systems, where the fluctuations between the regions
are reduced rather then enhanced. In line with this argument, the CSI criterion does not detect entanglement of the NOON state $\frac1{\sqrt2}(\ket{N,0}+\ket{0,N})$, which has maximal
fluctuations between the modes, i.e. $\eta^2=N$, while $\mathcal{C}=0$.

The CSI criterion from Eq.~(\ref{csi}) applies to systems with fixed number of particles,
such as an array of trapped bosonic ions \cite{wineland} or pairs of photons post-selected from the parametric down conversion signal \cite{para2,kwiat}.
However, for large systems, the number of particles is usually hard to control and differs between experimental realizations so in this context it is relevant to
extend above considerations to cases when $N$ fluctuates.
In such case, in the absence of coherence between states with different $N$, the density matrix of a separable state reads
\begin{equation}
  \hat\rho=\sum_{N=0}^\infty p_N\!\!\int\!\!\mathcal{D}\phi\,\ket{\phi;N}\!\!\bra{\phi;N}\mathcal{P}_N(\phi),
\end{equation}
where $p_N$ is the probability for having $N$ particles in the system, while $\mathcal{P}_N(\phi)$ for each $N$ and satisfies condition (\ref{cond_N}).
Now the CSI involves the integrals (\ref{integrated}) averaged with $p_N$ thus $N(N-1)\mathcal{P}(\phi)$  is replaced by
\begin{equation}
  \left\langle\mathcal{P}_N(\phi)\right\rangle\equiv\sum_Np_NN(N-1)\mathcal{P}_N(\phi).
\end{equation}
By inspecting the above formula, we note that even if some $\mathcal{P}_N(\phi)$ is partially negative, the CSI can overlook
entanglement present in this $N$-particle sector as long as the averaged $\left\langle\mathcal{P}_N(\phi)\right\rangle$ is positive.
From the point of view of inequality (\ref{csi}), the separable part of $\mathcal{P}_N$ can overshadow the entangled component.

The CSI criterion can be compared with yet another method of detecting entanglement in many-body systems known from quantum metrology.
The Quantum Fisher information (QFI), denoted here by $F_Q$ \cite{braun}, provides a lower bound for the precision $\Delta\theta$ of the estimation of an unknown parameter
$\theta$ in a series of $m$ experiments, $\Delta\theta\geqslant\frac1{\sqrt m}\frac1{\sqrt{F_Q}}$.
The value of $F_Q$ is determined by the properties of the state $\hat\rho$ and the transformation, which introduced the dependence on $\theta$ in the system.
A particularly important case is when  $\theta$ is the relative phase between two modes of $\hat\rho$, imprinted by an interferometer represented by a unitary transformation $e^{-i\theta\vec n\cdot\hat{\vec J}}$.
Here $\vec n\cdot\hat{\vec J}$ is a product of a unit vector and the vector of angular momentum operators \cite{nota_J}. This interferometric transformation addresses each particle independently,
so it cannot entangle them, and $F_Q\leqslant N$ holds for all separable $\hat\rho$ \cite{pezze}. Consequently, all two-mode states for which $F_Q>N$ must be entangled.% \cite{pezze}.
Typically it is not even necessary to find $\Delta\theta$ to estimate the value of $F_Q$. Usually in the experiment, some quantity $\chi\leqslant F_Q$ is measured,
such as for instance the inverse of the spin-squeezing parameter \cite{wine,kita,ita1,esteve,mzi_vienna,app,gross,riedel,vule,chen}.
If $\chi>N$, then also $F_Q>N$ meaning that the system is particle-entangled.

Contrary to the interferometric criterion, the CSI from Eq.~(\ref{csi}) does not involve any assumptions about the modal structure of $\hat\rho$.
Moreover, the QFI is inevitably related to an interferometric transformation, so to experimentally confirm the entanglement one is usually bound to use this interferometer
\cite{smerzi}. On the other hand, the CSI criterion is not linked with any transformation, and is solely based on the measurement of the integrated correlation
function (\ref{integrated}).

However, the metrological approach is more powerful than the violation of the CSI, because it is sensitive to a wider spectrum of entangled states -- 
thanks to the freedom of choice of the interferometric apparatus.
This can be illustrated by considering the bare phase imprint, which is represented by the transformation with $\vec n=(0,0,1)^T$.
This interferometer fed by the NOON state gives $F_Q=N^2$, while the CSI criterion does not detect entanglement, as argued above.

Another advantage of the QFI can be shown using a twin-Fock state $\ket{\psi_{\rm tf}}$ introduced above, which is passing through the Mach-Zehnder interferometer represented
by the interferometric transformation with $\vec n=(0,1,0)^T$. Then $F_Q=N+\frac{N^2}2$ showing that the state is strongly entangled and the correction
to the no-entanglement limit $F_Q=N$ is quadratic in $N$. In contrast the CSI gives  $\mathcal{C}=1+\frac2{N-2}$ so the deviation from the classical limit $\mathcal{C}=1$ becomes negligible for large $N$.

The above example underlines the main difference between the QFI approach and the CSI criterion.
The former, although usually difficult to implement, exploits information about the whole density matrix. The latter, much easier to check experimentally,
is based solely on the second-order correlation function.
Therefore, when $N$ is large, much knowledge about the non-classical relations between the particles, contained in higher order correlations, is lost.

One can overcome this limitation using the CSI calculated with higher order integrated correlation functions $\mathcal{G}^{(n)}_{ij}$ between $n/2$ particles in region $i$ and other $n/2$ in $j$.
When $N$ is large and $n\ll N$, 
the coefficient $\mathcal C$ for a twin-Fock state is $\mathcal{C}\simeq1+\frac12\frac{n^2}{N}$, showing increasing deviation from unity with growing $n$. This way it is possible to
increase the accuracy of entanglement detection for instance in cold-atoms systems, where ultra-precise measurements of spatial correlations up to sixth order were recently reported \cite{aust}.

Finally, we provide a simple example to illustrate that the indistinguishability of particles is crucial for the violation of the CSI to be used as an entanglement probe, while a rigorous discussion is
presented in the Supplementary Material.
Consider two particles occupying two modes $\phi^{(a)}$ and $\phi^{(b)}$ in a Werner state \cite{wer}
\begin{equation}
  \hat\rho_{\rm w}=\frac{1-p}4\hat{\mathds{1}}+p\ket{\psi_1}\bra{\psi_1},
\end{equation}
where $\ket{\psi_1}=\frac1{\sqrt2}(|\phi_1^{(a)},\phi_2^{(b)}\rangle+|\phi_2^{(a)},\phi_1^{(b)}\rangle)$ and $0\leqslant p\leqslant1$. Since the identity operator is spanned by the triplet of  bosonic vectors
$\ket{\psi_1}$, $\ket{\psi_2}=|\phi_1^{(a)},\phi_2^{(a)}\rangle$, $\ket{\psi_3}=|\phi_1^{(b)},\phi_2^{(b)}\rangle$ and a fermionic singlet
$\ket{\psi_4}=\frac1{\sqrt2}(|\phi_1^{(a)},\phi_2^{(b)}\rangle-|\phi_2^{(a)},\phi_1^{(b)}\rangle)$, then $\hat\rho_{\rm w}$ is not a state of indistinguishable bosons apart from $p=1$. For this state, the second
order correlation function can be easily calculated. For instance, $\mathcal{G}^{(2)}_{aa}=\tr{\hat\rho_{\rm w}\ket{\psi_2}\bra{\psi_2}}=\frac{1-p}4$ and symmetrically
$\mathcal{G}^{(2)}_{bb}=\frac{1-p}4$ while $\mathcal{G}^{(2)}_{ab}=\frac{1+p}2$, giving $\mathcal{C}=2\frac{1+p}{1-p}>1$ for all $0\leqslant p\leqslant1$. However, according the Peres-Horodecki positive partial
transpose criterion \cite{peres,horodecki}, $\hat\rho_{\rm w}$ is entangled only when $p>\frac13$, which confirms that indeed, the violation of the CSI does not imply entanglement of distinguishable particles.

In conclusion, we have demonstrated that the analog of the Glauber-Sudarshan P-representation used in quantum optics, can be employed to show that the violation of the CSI
for the second-order correlation function proofs entanglement in any system of identical bosons. 
The CSI condition is not as powerful as those inferred
from quantum metrology, but usually is much simpler to implement. 
It could be used as a direct test of entanglement for instance in systems, where the number-squeezing is likely to be present. Among those are the twin-beam setups \cite{bucker,lattice}, 
halos of particles scattered in the BEC collisions
\cite{Perrin07,jaskula,dall,pertot,rugway} and many other correlated systems. After the CSI violation is demonstrated, the systems are ready for potential applications, for instance 
in ultra-precise metrology.
%, measurements of Bell inequalities \cite{kwiat,ent2,Reid09} or demonstration of the EPR phenomenon.
The violation of the CSI was already measured in a collection of $^4$He particles emitted from a pair of colliding Bose-Einstein condensates (BECs) \cite{palais}.
The BECs are formed of identical bosons and since typical energies of the decohering processes are not high enough to distinguish the particles by changing their internal structure,
we conclude that this experiment demonstrated entanglement in a many-body system.

We ackowledge fruitful discussions with Denis Boiron and Andrew Truscott. 
T.W. and P. Sz. acknowledge the Foundation for Polish Science International Ph.D. Projects Programme co-financed by the EU European Regional Development Fund.
M. T. acknowledges the support of the National Science Center grant  N202 167840.
%J. Ch. acknowledges the Foundation for Polish Science International TEAM Programme co-financed by the EU European Regional Development Fund.
T. W., P. Sz., P. Z. and J.Ch. were supported by the National Science Center grant no. DEC-2011/03/D/ST2/00200.

\bigskip{}
\bigskip{}

\begin{center}
  {\bf Appendix}
\end{center}

\bigskip{}
\bigskip{}

\setcounter{equation}{0}
\makeatletter
\renewcommand{\theequation}{S\@arabic\c@equation}
\makeatother

In these Supplementary Materials, we derive a general expression for a separable state of $N$ identical bosons and then discuss the importance of the indistinguishability assumption for the CSI criterion.

\section{Appendix 1: Separable state of $N$ identical bosons}

Here we prove that the separable states of $N$ identical bosons has a form of Eq.~(6) as refered to in the article.
The calculation starts from the general form of the density matrix of a separable state of $N$ particles, which is a mixture of $N$-particle product states, i.e.
\begin{equation}\label{sep}
  \hat\rho = \sum_i p_i\, \hat\rho^{(1)}_i\otimes \cdots \otimes \hat\rho^{(N)}_i.
\end{equation}
Here, $\hat\rho^{(j)}_i$ is the density matrix of the $j$-th particle, while $p_i$ are the statistical weights of the mixture.
This $N$-body density matrix can be rewritten as
\begin{equation}\label{rho}
  \hat\rho =\sum_i P_i \proj{\psi_i},
\end{equation}
where each $N$-particle ket $\ket{\psi_i}$ is a product of $N$ single-particle pure states
\begin{equation}\label{sepket}
  \ket{\psi_i} = \ket{\phi_i^{(1)}} \otimes  \cdots \otimes \ket{\phi_i^{(N)}}.
\end{equation}
For indistinguishable bosons, only states, which are symmetrized with respect to the particle interchange, are permitted.
These states span the bosonic subspace $\mathcal{H}_B$ of full $N$-body Hilbert space $\mathcal{H}$.
We introduce the operator $\hat\Pi_B$, which projects onto $\mathcal{H}_B$ and note that if $\hat\rho$ describes a separable state of $N$ identical bosons, it must be unaltered by the
action of $\hat\Pi_B$.
In particular the equality
\begin{equation}
  \tr{\hat\rho} = \tr{\hat\Pi_B \hat\rho  \hat\Pi_B} =  \sum_i P_i \tr(\hat\Pi_B \proj{\psi_i})
\end{equation}
is fulfilled if $\tr{\hat\Pi_B \proj{\psi_n}} = 1$ for all  $i$, meaning that each $\ket{\psi_i}$ belongs to $\mathcal{H}_B$. The only pure state, which is symmetic and separable is a product
of $N$ identical single-particle states $\ket{\phi_i}$, so Eq.~(\ref{sepket}) simplifies to
\begin{equation}
  \ket{\psi_i} = \ket{\phi_i}^{\otimes N}\equiv \ket{\phi_i;N}.
\end{equation}
The general separable state of indistinguishable bosons is a mixture of such states and reads
\begin{equation}\label{disc}
  \hat\rho = \sum_i P_i \proj{\phi_i;N}
\end{equation}
or when the set of states $\ket{\phi_i;N}$ is continous
\begin{equation}\label{den}
  \hat\rho=\int\!\!\mathcal{D}\phi\,\ket{\phi;N}\!\!\bra{\phi;N}\mathcal{P}(\phi),
\end{equation}
where symbol $\mathcal{D}\phi$ denotes the measure of the integration over the set of states $\ket{\phi;N}$. This expression coincides with Eq.~(6) of the article.

\bigskip{}

{\bf Example: two qubits.}
In this example we apply the above formalism to determine the general form of the separable state of two identical qubits.

Consider a separable state of two qubits
\begin{equation}\label{2q}
  \hat\rho = \sum p_j \hat\rho_j^{(1)} \otimes \hat\rho_j^{(2)},
\end{equation}
where the density matrix of each qubit can be represented using the set of Pauli matrices $\vec{\hat\sigma}=(\hat\sigma_x,\hat\sigma_y,\hat\sigma_z)^T$
\begin{equation}
  \hat\rho_j^{(i)} = \frac12 \left(\hat 1 + \vec s_j^{\ (i)} \cdot \vec{\hat\sigma}^{(i)}\right).
\end{equation}
This operator is a valid density matrix when the length of the vectors $\vec s_j^{\ (i)}$ satisfies $|\vec s_j^{\ (i)}|\leqslant1$.

We introduce the triplet of bosonic states
\begin{subequations}
  \begin{eqnarray}
    &&\ket{\psi_1} = \ket{\uparrow}^{(1)}\otimes\ket{\uparrow}^{(2)}\\
    &&\ket{\psi_2} = \ket{\downarrow}^{(1)}\otimes\ket{\downarrow}^{(2)}\\
    &&\ket{\psi_3} = \frac{1}{\sqrt{2}}\left(\ket{\uparrow}^{(1)}\otimes\ket{\downarrow}^{(2)}+\ket{\downarrow}^{(1)}\otimes\ket{\uparrow}^{(2)}\right).
  \end{eqnarray}
\end{subequations}
where $\hat\sigma_z^{(i)}\ket{\uparrow\downarrow}^{(i)}=\pm\ket{\uparrow\downarrow}^{(i)}$ together with the operator
\begin{equation}
  \hat\Pi_B = \proj{\psi_1} + \proj{\psi_2} + \proj{\psi_3},
\end{equation}
which projects onto $\mathcal{H}_B$. The trace of the density matrix (\ref{2q}) projected onto $\mathcal{H}_B$ is
\begin{equation}
  \tr{\hat\Pi_B \hat\rho}=\frac14+\frac34\sum_{j} p_j\vec s_j^{\ (1)}\!\!\cdot\vec s_j^{\ (2)}.
\end{equation}
It is equal to unity only when $\vec s_j^{\ (1)}\!\! \cdot\! \vec s_j^{\ (2)} = 1$ for each $j$,
which means that $\vec s_j^{\ (1)} = \vec s_j^{\ (2)} \equiv \vec s_j$ and $|\vec s_j|^2=1$.
Therefore the states $\hat\rho_j^{(i)}$ are pure and identical for both qubits, which means that the bosonic density matrix reads
\begin{equation}
  \hat\rho = \sum p_j\left(\proj{\vec s_j}\right)^{\otimes2},
\end{equation}
which is consistent with Eq.~(\ref{disc}).

\section{Appendix 2: Importance of the indistinguishability assumption}

In this section we show that the indistinguishability of particles is vital for the violation of the CSI to be used as an entanglement criterion.
To this end, we calculate the second-order correlation function for a separable state (\ref{sep}) without imposing the indistinguishability of particles
\begin{equation}
  \mathcal{G}^{(2)}(x,x')=\sum_{n\neq m=1}^N\!\!\!\tr{\hat\rho_{\rm sep}\,\hat\Pi^{(n)}(x)\otimes\hat\Pi^{(m)}(x')}.
\end{equation}
Here, $\hat\Pi^{(n)}(x)$ projects the $n$-th particle onto the position state $\ket x$, while the sum ensures that all possible combinations of one particle being at position $x$ and the other
at $x'$ contribute to the correlation function. Using Eq.~(\ref{sep}) we obtain that
\begin{equation}
  \mathcal{G}^{(2)}(x,x')=\sum_ip_i\!\!\!\!\sum_{n\neq m=1}^N\!\!\!P_i^{(n)}(x)P_i^{(m)}(x'),
\end{equation}
where the one-body probability for finding the $n$-th particle in state $\hat\rho_i^{(n)}$ at position $x$ reads
\begin{equation}
  P_i^{(n)}(x)=\tr{\hat\rho_i^{(n)}\hat\Pi^{(n)}(x)}.
\end{equation}
If particles are identical, then these probabilities
do not depend  on indices $n$ and $m$.
In this case, the sum over $n\neq m$ gives the coefficient $N(N-1)$ and after integrating $x$ over volume $V_i$ and $x'$ over $V_j$, we obtain the discrete version of Eq.~(9).
However, if particles are not identical, then the sum over $n\neq m$ gives
\begin{equation}\label{dist}
  \mathcal{G}^{(2)}(x,x')=\sum_ip_if_i(x,x'),
\end{equation}
where $f_i(x,x')$ does not factorize into a product of functions of $x$ and $x'$. In consequence, after the integration of Eq.~(\ref{dist}) over the regions $a$ and $b$, it is not possible to introduce
two separate functions $I_a(\phi)$ and $I_b(\phi)$. Therefore, no such relation as in Eq.~(11) of the article can be established.


\begin{thebibliography}{40}

\bibitem{sud} E. C. G. Sudarshan, \PRL{10}, 277 (1963)
\bibitem{glaub} R. J. Glauber, Phys. Rev. {\bf 131}, 2766 (1963)

\bibitem{ent1} I. Bengtsson and K. \.Zyczkowski, {\it Geometry of Quantum States. An Introduction to Quantum Entanglement}, Cambridge University Press (2006)
\bibitem{ent2} R. Horodecki, P. Horodecki, M. Horodecki and K. Horodecki, Rev. Mod. Phys. {\bf 81}, 865 (2009)

\bibitem{epr} A. Einstein, B. Podolsky, and N. Rosen, Phys. Rev. {\bf 47}, 777 (1935)

\bibitem{quantinf} M. A. Nielsen and I. L. Chuang, {\it  Quantum computation and quantum information}, Cambridge University Press (2000).

\bibitem{wootters1} C. H. Bennett, G. Brassard, C. Cr\'epeau, R. Jozsa, A. Peres, and W. K. Wootters, Phys. Rev. Lett. {\bf 70}, 1895 (1993)
\bibitem{wootters2} C. H. Bennett, G. Brassard, S. Popescu, B. Schumacher, J. A. Smolin, and W. K. Wootters, Phys. Rev. Lett. {\bf 76}, 722 (1996)

\bibitem{giovanetti} {V. Giovanetti, S. Lloyd and L. Maccone, Science {\bf 306}, 1330 (2004)}
\bibitem{pezze}{L. Pezz\'e and A. Smerzi, Phys. Rev. Lett. {\bf 102}, 100401 (2009)}

\bibitem{synt} M. Sarovar, A. Ishizaki,	G. R. Fleming and K. B. Whaley, Nat. Phys. {\bf 6}, 462 (2010)

\bibitem{decoh1} M. Schlosshauer, {\it Decoherence and the Quantum-to-Classical Transition}, Springer (2007)
\bibitem{decoh2} E. Joos, H. D. Zeh, C. Kiefer, D.J.W. Giulini, J. Kupsch and I.-O. Stamatescu, {\it Decoherence and the Appearance of a Classical World in Quantum Theory}, Springer (2003)
\bibitem{decoh3} R. Omnes, {\it  Understanding Quantum Mechanics}, Princeton University Press (1999)
\bibitem{decoh4} W. H. Zurek, arXiv:quant-ph/0306072
\bibitem{decoh5} M. Schlosshauer, Rev. Mod. Phys. {\bf 76}, 1267 (2004)

\bibitem{sor}{A. Sorensen, L.-M. Duan, J.I. Cirac and P. Zoller, Nature {\bf 69}, 63 (2001)}
\bibitem{peres} A. Peres, \PRL{77}, 1413 (1996)
\bibitem{wer} R. F. Werner, Phys. Rev. A {\bf 40}, 4277 (1989)

\bibitem{sym1} M. Nielsen and I. Chuang, {\it Quantum Computation and Quantum Information}, Cambridge University Press (2000)
\bibitem{sym2} A. Peres, {\it Quantum Theory: Concepts and Methods}, Kluwer (1995)
\bibitem{sym3} P. Zanardi, Phys. Rev. A {\bf 65}, 042101 (2002)
\bibitem{sym4} F. Benatti, R. Floreanini and U. Marzolino, Ann. Phys. {\bf 325}, 924 (2010)

\bibitem{plenio} N. Killoran, M. Cramer and M. B. Plenio, arxiv:1312.4311v1

\bibitem{wine} { D. J. Wineland, J. J. Bollinger, W. M. Itano and D. J. Heinzen, \PRA{50}, 67 (1994)}
\bibitem{kita} {M. Kitagawa and M. Ueda, \PRA{47}, 5138 (1993)}
\bibitem{ita1}{D. J. Wineland, J. J. Bollinger, W. M. Itano, F. L. Moore and D. J. Heinzen, \PRA{46}, 6797 (1992)}

\bibitem{app}{J. Appel, P. J. Windpassinger, D. Oblak, U. B. Hoff, N. Kj\ae rgaard, and E. S. Polzik, PNAS {\bf 106}, 10960 (2009)}
\bibitem{gross}{C. Gross, T. Zibold, E. Nicklas, J. Esteve and M. K. Oberthaler, Nature {\bf 464}, 1165 (2010)}
\bibitem{riedel}{M. F. Riedel, P. Bohi, Y. Li, T. W. Hansch, A. Sinatra and P. Treutlein, Nature {\bf 464}, 1170 (2010)}
\bibitem{vule}{I. D. Leroux, M. H. Schleier-Smith and V. Vuletic, \PRL{104}, 1170 (2010)}
\bibitem{chen}{Z. Chen, J. G. Bohnet, S. R. Sankar, J. Dai, and J. K. Thompson, \PRL{106}, 133601 (2011)}
\bibitem{esteve}{J. Est\'eve, C. Gross, A. Weller, S. Giovanazzi and M. K. Oberthaler, Nature {\bf 455}, 1216 (2008)}
\bibitem{mzi_vienna}{T. Berrada, S. van Frank, R. B\"ucker, T. Schumm, J.-F. Schaff and J. Schmiedmayer, Nat. Comm. {\bf 4}, 2077 (2013)}

\bibitem{proof1} T. Ichikawa, T. Sasaki, I. Tsutsui and N. Yonezawa, Phys. Rev. A {\bf 78}, 052105 (2008)
\bibitem{proof2} Tzu-Chieh Wei, Phys. Rev. A {\bf 81}, 054102 (2010)


\bibitem{shch} E. Shchukin and W. Vogel, Phys. Rev. Lett. {\bf 95}, 230502 (2005)
\bibitem{hillery} M. Hillery and M. S. Zubary, Phys. Rev. Lett. {\bf 96}, 050503 (2006)
\bibitem{tests} A. Miranowicz, M. Bartkowiak, X. Wang, Y. Liu and F. Nori, Phys. Rev. A {\bf 82}, 013824 (2010)

\bibitem{wodk} B.-G. Englert and K. W\'odkiewicz, Phys. Rev. A {\bf 65}, 054303 (2002)

\bibitem{wineland} D. Leibfried, R. Blatt, C. Monroe, and D. Wineland, Rev. Mod. Phys. {\bf 75}, 281 (2003)
\bibitem{para2} D. C. Burnham and D. L. Weinberg, Phys. Rev. Lett. {\bf 25}, 84 (1970)
\bibitem{kwiat} P. G. Kwiat, K. Mattle, H. Weinfurter, A. Zeilinger, A. V. Sergienko and Y. Shih, Phys. Rev. Lett. {\bf 75}, 4337 (1995)


\bibitem{braun} S. L. Braunstein and C. M. Caves, Phys. Rev. Lett. {\bf 72}, 3439 (1994).

\bibitem{nota_J}{
  Angular momentum operators are $\hat J_x \equiv (\hat a^\dagger\hat b^{\phantom{\dagger}}+
  \hat b^\dagger\hat a^{\phantom{\dagger}})/2$, $\hat J_y \equiv (\hat a^\dagger\hat b^{\phantom{\dagger}}-
  \hat b^\dagger\hat a^{\phantom{\dagger}})/2i$ and $\hat J_z \equiv (\hat a^\dagger\hat a^{\phantom{\dagger}}-
  \hat b^\dagger\hat b^{\phantom{\dagger}})/2$.}

\bibitem{smerzi} B. L\"ucke,  M. Scherer,  J. Kruse,  L. Pezz\,e, F. Deuretzbacher, P. Hyllus, O. Topic, J. Peise, W. Ertmer, J. Arlt, L. Santos, A. Smerzi and C. Klempt,
  Science {\bf 11}, 773 (2011)

\bibitem{aust} R. G. Dall, A. G. Manning, S. S. Hodgman, Wu RuGway, K. V. Kheruntsyan and A. G. Truscott, Nat. Phys. {\bf 9}, 341 (2013)

\bibitem{horodecki} M. Horodecki, P. Horodecki and R. Horodecki, Phys. Lett. A {\bf 223}, 1 (1996)

\bibitem{bucker} R. B\"ucker, J. Grond, S. Manz, T. Berrada, T. Betz, C. Koller, U. Hohenester, T. Schumm, A. Perrin and J. Schmiedmayer, Nat. Phys. {\bf 7}, 608 (2011)
\bibitem{lattice} M. Bonneau, J. Ruaudel, R. Lopes, J.-C. Jaskula, A. Aspect, D. Boiron, and C. I. Westbrook, Phys. Rev. A {\bf 87}, 061603 (2013)

\bibitem{Perrin07} A.~Perrin, H.~Chang, V.~Krachmalnicoff, M.~Schellekens, D.~Boiron, A.~Aspect, C.I.~Westbrook, \PRL{99}, 150405 (2007).
\bibitem{jaskula} J.-C. Jaskula, M. Bonneau, G. B. Partridge, V. Krachmalnicoff, P. Deuar, K. V. Kheruntsyan, A. Aspect, D. Boiron, and C. I. Westbrook, Phys. Rev. Lett. {\bf 105}, 190402 (2010)
\bibitem{dall} R. G. Dall, L. J. Byron, A. G. Truscott, G. R. Dennis, M. T. Johnsson, and J. J. Hope, Phys. Rev. A {\bf 79}, 011601 (2009)
\bibitem{pertot} D. Pertot, B. Gadway, and D. Schneble, Phys. Rev. Lett. {\bf 104}, 200402 (2010)
\bibitem{rugway} Wu RuGway, S. S. Hodgman, R. G. Dall, M. T. Johnsson, and A. G. Truscott, Phys. Rev. Lett. {\bf 107}, 075301 (2011)


\bibitem{Reid09} M. D. Reid, P. D. Drummond, W.P. Bowen, E. G. Cavalcanti, P. H. Lam, H. A. Bachor, U. L. Andersen, G. Leuchs, Rev.~Mod.~Phys.~\textbf{81}, 1727 (2009).
\bibitem{palais} K. V. Kheruntsyan, J.-C. Jaskula, P. Deuar, M. Bonneau, G. B. Partridge, J. Ruaudel, R. Lopes, D. Boiron, and C. I. Westbrook, Phys. Rev. Lett. {\bf 108}, 260401 (2012)

\end{thebibliography}
\end{document}